\documentclass{aastex}
\usepackage{emulateapj5,apjfonts,psfig}

\makeatletter
\newenvironment{figurehere}
  {\def\@captype{figure}}
  {}
\makeatother

\newcommand\fu{{4U~2129+47}}

\newcommand\axaf{\textsl{Chandra}}
\newcommand\chandra{\textsl{Chandra}}

\newcommand\hst{\textsl{HST}}
\newcommand\fos{\textsl{FOS}}

\newcommand\rosat{\textsl{ROSAT}}
\newcommand\hri{\textsl{HRI}}
\newcommand\pspc{\textsl{PSPC}}

\newcommand\xmm{\textsl{XMM-Newton}}
\newcommand\msun{{\rm M_\odot}}
\newcommand\aproxgt{\mathrel{%
      \rlap{\raise 0.511ex \hbox{$>$}}{\lower 0.511ex \hbox{$\sim$}}}}
\newcommand\aproxlt{\mathrel{%
      \rlap{\raise 0.511ex \hbox{$<$}}{\lower 0.511ex \hbox{$\sim$}}}}
\def\errtwo#1#2#3{$#1^{+#2}_{-#3}$}

\begin{document}

\slugcomment{Submitted 2001 December 1; Accepted 2002 March 19}

\shortauthors{Nowak, Heinz, \& Begelman}
\shorttitle{Total Eclipses of \fu}

\title{Hiding in Plain Sight: \axaf\ Observations of the Quiescent Neutron
  Star \fu\ in Eclipse}
\author{Michael A. Nowak} 
\affil{MIT, Center for Space Research, NE80-6077, 77 Massachusetts Ave., Cambridge,
  MA 02139, U.S.A.}
\email{mnowak@alum.mit.edu}

\author{Sebastian Heinz} 
\affil{Max-Planck-Institut fuer Atrophysik, Karl-Schwarzschild-Str. 1,
  85740 Garching bei Muenchen} 
\email{heinz@MPA-Garching.MPG.DE}

\author{M.C. Begelman}
\affil{JILA and Dept. of APS, University of Colorado, Boulder, CO 80309}
\email{mitch@jila.colorado.edu}

\begin{abstract} During a previous outburst phase, the neutron star \fu\
  exhibited evidence for a spatially extended Accretion Disk Corona
  (ADC) via broad, partial X-ray eclipses occurring periodically on the
  5.24\,hr orbit. Since 1983, however, \fu\ has entered a quiescent
  state several orders of magnitude fainter.  We have performed a
  37\,ksec ($\approx$2 binary orbits) \chandra\ observation of \fu\ to
  determine whether an extended coronal structure also exists in
  quiescence.  Total eclipses are found, and the rapidity of the
  eclipse ingress and egress are used to place upper limits on the
  size of the X-ray source.  The spectrum is found to be comprised of
  a soft component plus a fairly hard power law tail.  The former is
  seen to be sinusoidally modulated over the orbital period in a
  manner consistent with neutral column variations, possibly due to
  the interaction of an accretion stream with a (small) disk about the
  neutron star.  We fit realistic neutron star atmosphere models to
  the soft spectra, and comment on the consistency of the spectra with
  cooling neutron star models.  It has been previously suggested that
  the \fu\ system is a hierarchical triple, with the outer body being
  an F star.  We use differential astrometry to show that the X-ray
  point source and F star are spatially coincident to within
  $0.1\arcsec$.  We further compare newly determined upper limits for
  the extrinsic neutral column to the reddening of the F star.
  Finally, we discuss how future X-ray observations can further
  constrain models of quiescent neutron star emission, as well as
  directly verify the triple hypothesis.
\end{abstract}


\section{Introduction}\label{sec:intro}

`Accretion Disk Corona' (ADC) sources such as \fu\ are believed to be
near edge-on accreting systems since they have shown binary orbital
modulation via broad, partial X-ray eclipses.  Such eclipses imply
emission extended over radii comparable to the binary separation
\citep{mcclintock:82a,white:82a}.  Studies of the ADC X1822$-$371
offer an intriguing possibility for the cause of this very extended
corona \citep{heinz:01a}.  Orbital period derivative measurements
imply that X1822$-$371 may be accreting very close to its Eddington
rate; however, only a very small fraction of the Eddington flux is
observed.  If the mass transfer from the secondary is actually
accreted by the primary, then the ADC must either be very optically
thin (the tacit assumption made for `on state' emission of \fu\ by
Garcia \& Grindlay \nocite{garcia:87a} 1987), or very optically thick.
If the latter is true, the released accretion energy may not be able
to diffuse out of the flow on a viscous inflow time scale, and the
corona will essentially become an `advection dominated accretion flow'
(ADAF; Esin, Narayan, \& McClintock \nocite{esin:97c} 1997, and
references therein). In order to dissipate the accretion energy, an
adiabatic outflow might ensue at large radius (`Adiabatic
Inflow/Outflow Solution', ADIOS; Blandford \& Begelman
\nocite{blandford:99a} 1999), hence creating the ADC.

The ADC source \fu\ is strongly believed to be a neutron star since it
has exhibited a Type I X-ray burst \citep{garcia:87a}.  Its
classification as an ADC is due to the fact that prior to 1983, \fu\ 
was observed to have a lightcurve similar to that of X1822$-$371.
Over the 5.24\,hr orbit of \fu, the eclipse width was $\approx 0.2$ in
phase and $\approx 75$\% of the X-rays were occulted at the eclipse
midpoint.  As hypothesized for X1822$-$371, the ADC might have been
indicating the presence of an ADIOS.

Since 1983, \fu\ has been in a quiescent state.  The question then
arises as to whether the extended coronal structure has persisted or
recurred in this state (see Garcia \nocite{garcia:94a} 1994; Garcia \&
Callanan \nocite{garcia:99a} 1999).  An extended coronal structure
could signal the presence of an ADIOS, albeit at low accretion rates
where advection domination might ensue due to a decoupling of the ions
and electrons.  The bulk of the accretion energy would stored as
thermal energy of the protons, which would then be lost as a wind
rather than be dissipated on the neutron star surface
\citep{blandford:99a}.

Alternatively, the spectra of quiescent neutron star sources might be
due emission from their surface as they cool after a prior active
accretion phase, or due to a low level of residual accretion
\citep{vanparadijs:87a,brown:98a,rutledge:00a,bildsten:01a}.  Thermal
emission from the neutron star surface has been suggested as partly
contributing to the putative differences between the fluxes of
quiescent neutron stars and quiescent black hole candidates, with the
former class being suggested as systematically brighter
\citep{narayan:97b,bildsten:01a,garcia:01a}.  Modern neutron star
atmosphere models utilizing realistic opacities
\citep{raj:96a,zavlin:96a} have yielded good fits to \rosat\ spectra
of quiescent neutron stars, including quiescent spectra of \fu\ 
\citep{rutledge:00a}.  The flux and color temperature of the spectrum
are determined by the level of residual accretion and/or by the time
averaged (over $10^4$--$10^5$\,yr time scales) accretion rate in prior
active phases \citep{brown:98a}.  The degree to which residual heat
dominates over active accretion in specific sources is a matter of
current debate, with there being a recent suggestion that the
quiescent flux level of \fu\ might be too low to be consistent with
the cooling model \citep{wijnands:02a}.

It is also possible that for quiescent neutron stars the ``propeller
effect'' (Campana et al. \nocite{campana:98a} 1998, and references
therein) halts, or at least greatly diminishes, the accretion rate
onto the neutron star surface.  The neutron star might then enter a
radio pulsar phase, with X-ray emission arising from the interaction
between the pulsar wind and the accretion flow from the secondary
\citep{campana:98a,menou:99a,campana:00a}.  The X-ray emission would
be prominent at very large radii, comparable to the disk
circularization radius.  As in the outburst phase of \fu, this would
imply that the X-ray emission would be gradually eclipsed over
$\aproxgt 0.1$ in orbital phase.

A number of neutron star systems in quiescence have exhibited hard
tails \citep{asai:96a,asai:98a,campana:98a,rutledge:01a}, which might
be related to the shocked pulsar wind.  The shocked wind emission
might be in addition to and somewhat independent of any emission
associated with the neutron star surface, i.e., the hard tail could
vary while the soft excess remained steady \citep{campana:00a}.  The
presence or absence of such a hard tail in the quiescent state of \fu\ 
has been impossible to determine with \rosat\ observations
(0.1-2.4\,keV), although those observations did suggest that the soft
X-ray component was variable \citep{garcia:94a,garcia:99a}.

\fu\ has another very interesting property which relates to estimates
of its quiescent flux.  After \fu\ entered quiescence, optical
modulation was no longer observed over its 5.24\,hr binary period.
Instead of finding its expected M or K type companion, a nearby late F
type star was found
\citep{kaluzny:88a,thorstensen:88a,chevalier:89a,garcia:89a,cowley:90a},
which has led to the hypothesis that the \fu\ system is a hierarchical
triple.  The F star would be in a few month orbit about the inner
binary, which in turn could lead to modulations of the binary
eccentricity on 50\,yr time scales \citep{garcia:89a}.  It is the
optical measurements of this F star, which may or may not be related
to the X-ray source, that yields the 6.3\,kpc distance that has been
used by \citet{garcia:01a} to determine the quiescent flux of \fu.  We
note that given a 6.3\,kpc distance, \fu\ is the third brightest
quiescent neutron star in the \citet{garcia:01a} sample.

The \chandra\ X-ray satellite, which provides excellent spatial
resolution ($\aproxlt 1\arcsec$) and good coverage in the soft X-ray
band (0.5--10\,keV) can help answer some of the questions concerning
the \fu\ system.  The high spatial resolution can determine whether or
not the X-ray source is truly coincident with the observed F star.
Measurement of the eclipse can determine whether an extended coronal
structure still exists and place constraints on the mass and radius of
the secondary.  The 0.5--2\,keV spectra can yield constraints on
surface emission models, while the 2--10\,keV spectra can indicate
whether \fu\ exhibits a hard tail similar to other quiescent neutron
stars. To these ends, on 9 December 2000, we obtained a 37\,ksec
($\approx2$ binary orbital periods) observation of \fu.  Details of
the observational modes and data extraction methods are presented in
Appendix~\ref{sec:extract}.

The discussion of our results is organized as follows.  We first
outline our data extraction methodology in \S\ref{sec:extract}.  In
\S\ref{sec:astrometry} we discuss the previous optical observations
and the extent to which the X-ray source is coincident with the
identified F star.  We then present the X-ray lightcurve in
\S\ref{sec:curve} and discuss its implications.  We present orbital
phase resolved spectroscopy in \S\ref{sec:spec}.  We summarize our
results in \S\ref{sec:summary}, and suggest further observations with
the \textsl{X-ray Multi-Mirror Mission (XMM)} that can resolve the
mysteries left unanswered by these \chandra\ observations.  So as not
to deter the reader from the scientific results, some of the details
of the analyses are relegated to the Appendix, including discussion of
the statistical methods used in analyzing the lightcurve
(Appendix~\ref{sec:anal}).  In Appendix~\ref{sec:serendipity}, we
briefly discuss the sources that were serendipitously observed along
with \fu.

\section{Data Analysis Methodology}\label{sec:extract}

All observations of \fu\ described here were taken with the
\textsl{Advanced Camera from Imaging Spectroscopy} (\textsl{ACIS}) on
board the \chandra\ X-ray satellite.  Specifically, they were taken
with the backside illuminated S3 chip (in imaging mode, i.e., without
the gratings inserted).  To reduce the amount of spectral pile-up (see
below), we employed a 1/4 sub-array on all chips, yielding a frame
readout time of 1.1\,sec.  The total integration time was
approximately 36.6\,ksec, with roughly a 96.5\% observing efficiency
(see below), yielding 35.3\,ksec of on source time. Data extraction
and response matrix generation were performed with the {\tt CIAO v2.2}
analysis package and calibration products from {\tt CALDB v2.9}.
Sources within the field of view were detected using the {\tt
  wavdetect} program, and source counts were extracted from within the
$3\sigma$ contours found from this analysis.  The \fu\ background
region was extracted from an annulus centered on the source with an
inner radius of 5\arcsec\ and an outer radius of 14\arcsec.  We
include in our spectral \emph{and} lightcurve analyses photons only in
the 0.5--8\,keV range, and have rebinned the spectra to have between
20--26 photons per spectral bin.  In determining the error bars of the
spectral parameters, we have calculated the 90\% confidence limits
using the methods of \cite{lampton:76a} (i.e., $\Delta\chi^2=2.71$ for
one interesting parameter).  We have not applied any systematic errors
to the spectral energy bins.

Although the detected count rate from \fu\ was only $\aproxlt
0.05$\,cps, the observation did suffer from a small degree of pile-up
(i.e., two photons being detected as a single photon with their summed
energy).  An in depth discussion of pile-up, along with analysis
methods that may be able to account partially for its effects, is
presented by \citet{davis:01a}.  In all fits presented in this work,
we have employed the pile-up model of \citet{davis:01a} as implemented
in {\tt ISIS v1.0.6} \citep{houck:00a}.

With a 1.1\,sec frame integration time and an $\approx 0.05$\,cps
count rate, we expect 28 double photon events over the course of our
entire observation.  Roughly 95\% of the detected photons land on the
same $3\times3$ pixel region, which implies that 10\% of the double
photon events were registered as individual photons, whereas the
remaining 90\% of the double events were potentially registered as a
single photon.  The greatest uncertainty in this latter number is the
``grade migration factor'', $\alpha$, i.e., the probability that the
piled-up event is \emph{not} rejected as a cosmic ray particle or
other background event.  In our use of the {\tt ISIS} implementation
of the model of \citet{davis:01a}, we have assumed that there is a
50\% probability ($\alpha=0.5$) that two piled-up photons are accepted
as a single ``good event''.

Consistent with the above expectations, two lightcurve time bins show
two distinguished photons.  Since the detected spectrum peaks at
$\approx 1$\,keV, the majority of piled-up events accepted as photons
likely were read as having energy $\aproxgt 2$\,keV. Even though we
only expect $\aproxlt 26$ piled-up events, as we discuss below there
are only $\approx 100$ photons in the hard tail of our observed
spectrum; therefore, pile-up does effect the fitted power law slope
and amplitude.  If $\alpha=0.5$ is in fact too low a probability for
accepting a piled-up event, then we have under subtracted the piled-up
photons at 2\,keV and slightly overestimated the value of the photon
index.  Likewise, if $\alpha=0.5$ is too large, then we have slightly
underestimated the photon index, $\Gamma$.  To date, there has been no
independent calibration of the value of $\alpha$ (which in reality is
likely to be energy and chip location dependent) in the \chandra\ 
detectors. Our estimates are that for the fits described below these
systematic effects lead to at most $\Delta \Gamma \approx \pm 0.3$.

\section{Astrometry}\label{sec:astrometry}

Based upon the assumption of Roche lobe overflow and a main sequence
mass-radius relationship, the 5.24\,hr orbital period of \fu\ suggests
a secondary mass of $\approx 0.6\,\msun$, which corresponds to an M or
K dwarf.  It was expected that optical observations of the \fu\ system
in quiescence would reveal strong ellipsoidal variations.  Instead,
\citet{thorstensen:88a} observed a system exhibiting an F star with no
detectable orbital modulations.  Comparison of the position of this
quiescent optical source with the position of the optical counterpart
of \fu\ in outburst revealed the two sources to be coincident within
0.3\arcsec; therefore, \citet{thorstensen:88a} considered the
possibility that the \fu\ system might be a triple.  With higher
resolution optical observations, \citet{chevalier:89a} showed that the
\fu\ field was fairly crowded, but that there were no obvious
contaminating sources.  These authors favored the triple hypothesis.
\citet{garcia:89a} found that the F star exhibited spectroscopic
velocity modulations on the order of tens of km~s$^{-1}$ on the time
scale of months, which they also attributed to the system being a
triple.

In Table~\ref{tab:optical} we list some of the optical identifications
made for the quiescent counterpart to \fu.  All spectroscopic
observations, ranging from 2200--7500\,\AA, agree on some type of F
star identification for this counterpart, and most observations
suggest an extinction of $E(B-V) \approx 0.3$. If the X-ray source is
at the same distance as the F star, then it is at a distance of $\sim
4$--8\,kpc.  (The value of 6.3\,kpc used by Narayan, Garcia, \&
McClintock \nocite{narayan:97a} 1997 and Garcia et al.
\nocite{garcia:01a} 2001 is taken from Cowley \& Schmidtke
\nocite{cowley:90a} 1990.)  With the exquisite spatial resolution of
\chandra, we now can also determine to what degree the X-ray source is
coincident with the F star.  The J2000 X-ray source position---
RA$={\rm 21^h\,31^m\,26.19^s}$,
DEC$=47^\circ\,17\arcmin\,24.7\arcsec$--- agrees very well with the
coordinates of the F star observed by the \textsl{Hubble Space
  Telescope Faint Object Spectrograph} (\hst-\fos)
\citep{deutsch:96a}.  

The spatial coincidence can be further tested with the \hst\ images
obtained (separately from the \textsl{FOS} spectra) by
\citet{deutsch:96a}.  In addition to the image of the F star, the
\hst\ observations also contain a likely counterpart for the source
S3-$\delta$ listed in Appendix~\ref{sec:serendipity}.  The S3-$\delta$
X-ray source, likely a star, has 28 photons detected over the course
of the observation, and therefore is well-localized.  Applying an
$\approx 2$\arcsec\ translation\footnote{As described by threads on
  the \textsl{Chandra} and {\textsl HST} web pages ({\tt
    http://asc.harvard.edu} and {\tt http://www.stsci.edu},
  respectively), if one does not cross calibrate the acquired images
  with an accurate astrometric database, absolute astrometry errors of
  $0.5\arcsec$--$1.5\arcsec$ are expected \emph{for each instrument}.
  As we are here only interested in differential astrometry, we have
  not attempted to absolutely calibrate the coordinates of each image
  field.}, linear in RA and DEC, to center the X-ray image of \fu\ on
the \hst\ image of the F star, it is found that the X-ray image of the
S3-$\delta$ source coincides with an optical counterpart to within
0.1\arcsec.  This is near the limit for differential astrometry
performed with \chandra.  The \fu\ X-ray source and the optically
identified F star discussed in the references listed in
Table~\ref{tab:optical} are therefore, to good accuracy, spatially
coincident.  In \S\ref{sec:spec}, we comment further upon the extent
to which the F star and the quiescent X-ray counterpart of \fu\ are
at the same distance.

\vspace{0.2in}
\begin{figurehere}
\centerline{
\includegraphics[width=0.45\textwidth]{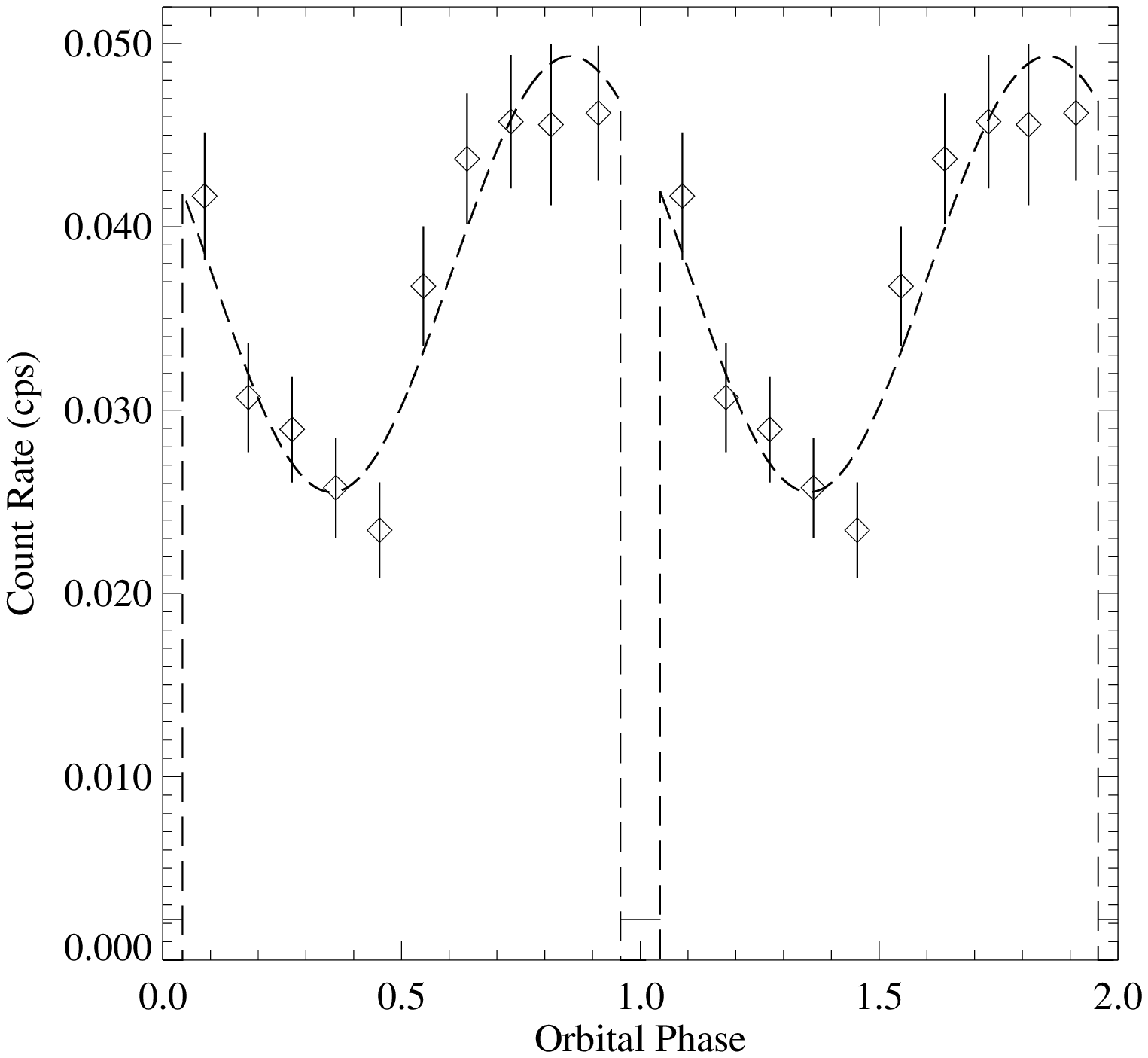}
}
\caption{\small Folded 0.5--8\,keV lightcurve of \fu, (repeated at phases
  1--2 for clarity).  Error bars are 1-$\sigma$ Poisson errors, except
  during the eclipse where the horizontal bar is the 3-$\sigma$ upper
  limit to the count rate.  The dashed line shows the best fitting
  sinusoidal modulation plus a total, rapid eclipse.
  \protect{\label{fig:lcurve}}}
\end{figurehere}

\section{Lightcurve}\label{sec:curve}

One of the main goals of our observation was to determine whether or
not the broad, partial dip seen in the outburst lightcurve of \fu\ 
\citep{mcclintock:82a,white:82a} persisted into quiescence.
\citet{garcia:94a} did not detect any evidence of an eclipse in
quiescence (see below), while \citet{garcia:99a} found only weak
evidence for an eclipse and could not distinguish among total and
partial eclipse models.  The folded lightcurve from our observation,
shown in Fig.~\ref{fig:lcurve}, removes any ambiguity and shows 
that the eclipse is rapid, total, and of shorter
duration than in outburst.  The eclipse duration is
$1523^{+30}_{-50}$\,sec (see Appendix~\ref{sec:anal}), i.e., 0.08 in
binary phase. This is 2.5 times narrower than in outburst. No photons
are detected over the course of the eclipse, which implies that the
3$\sigma$ upper limit to the mean observed count rate during eclipse
is $\approx 0.002$\,cps.  Given the mean count rate out of eclipse,
the eclipsing object exhibits an optical depth of $\tau \ge 3$ with
99.9\% confidence, and $\tau \ge 4$ with 95\% confidence.

In addition to the eclipse, the lightcurve exhibits a sinusoidal
modulation.  Fitting the lightcurve outside of eclipse with a function
$R(\phi) = A + B \sin [ 2 \pi (\phi - \phi_0) ]$, where $R$ is the
observed rate and $\phi$ is the orbital phase ($\phi \equiv 0$ is the
midpoint of the eclipse), we find $A = 0.037 \pm 0.002$, $B = 0.012
\pm 0.002$, and $\phi_0 = 0.60 \pm 0.03$ (90\% confidence levels).
Note that this sinusoidal modulation is similar to that exhibited by
X1822$-$371 in its bright state \citep[][and references
therein]{parmar:00a,heinz:01a}, where it is believed that modulation
is due to interactions of the accretion stream with the accretion
disk.  Thus the sinusoidal modulation may be indicating the presence
of an active accretion stream and a disk about the neutron star.

Note that \citet{garcia:94a} in three separate, short
\textsl{ROSAT-HRI} observations found evidence of variability.  Due to
the fact that these observations were spread over the span of a month,
one could argue whether intrinsic source variability or orbital
modulation was observed.  The sinusoidal modulation observed here is
roughly consistent, in both phase and amplitude, with the X-ray
variability described by \citet{garcia:94a} (who likely detected
little, if any, of the actual eclipse). Thus, whereas the quiescent
lightcurve of \fu\ has changed from that in outburst
\citep{garcia:94a}, it is possible that the quiescent lightcurve has
remained remarkably steady, at least in shape and \emph{relative}
amplitude, during quiescence.

With an accurate measurement of the eclipse, we can update the
ephemeris presented by 
\citet{mcclintock:82a}. The barycenter corrected Julian Date
of the first eclipse ($\phi=0$) in our lightcurve is
\begin{equation}
{\rm JD} ~=~ 2451879.5713 \pm 0.0002 ~~,
\end{equation}
which is 0.0293\,days later than expected from the 
\citet{mcclintock:82a} ephemeris.  This discrepancy could be due to
either the uncertainty in the original orbital period determination, or due
to orbital evolution.  Adding a quadratic term to the 
\citet{mcclintock:82a} ephemeris, and propagating errors, we find
the time of phase zero to be
\begin{eqnarray}
T_{0} = 2444403.7429\pm0.002 + (0.2182579\pm 8\times 10^{-7}) ~N \cr
   + [(5.0 \pm 2.4) \times 10^{-11}] ~N^2 ~.
\end{eqnarray}
(For these observations, we have observed orbits $N=34252$ and $34253$.)
Orbital evolution is indicated, but only at the 2$\sigma$ level.
Converting this into an orbital period derivative, we find
\begin{equation}
\dot P^{-1} P ~=~ (1.6\pm0.8) \times 10^6\, {\rm years} ~~.
\end{equation}

Due to the possibility of winds, magnetic torques, etc., it can
problematic to convert an observed period derivative directly into an
implied accretion rate \citep[see, for example,][for a general
discussion of the role of some of these effects]{accretion}.  However,
if the above period derivative is real and solely due to accretion
from the secondary to the compact object, the implied accretion rate
averaged over the past twenty years is $\langle \dot M \rangle =
(4\pm2) \times 10^{18}~{\rm g~s^{-1}}$.  Given that \fu\ has been in
quiescence over most of this time, this would imply an extremely large
accretion rate in outburst, and/or substantial mass loss without
radiation, i.e., an ADIOS, in quiescence.  Comparably large orbital
period derivatives and implied mean accretion rates have been found
for X1822$-$371 \citep{parmar:00a,heinz:01a}.

We can also determine an ephemeris for the \emph{average} period.
Specifically, we have
\begin{equation}
T_{0} = 2451879.5713\pm0.0002 + (0.21825962\pm 6\times 10^{-8}) ~N ~,
\end{equation}
where the period at the time of these observations might be as much as
$10^{-6}$\,days longer than the above average period.

\vspace{0.2in}
\begin{figurehere}
\centerline{
\includegraphics[width=0.45\textwidth]{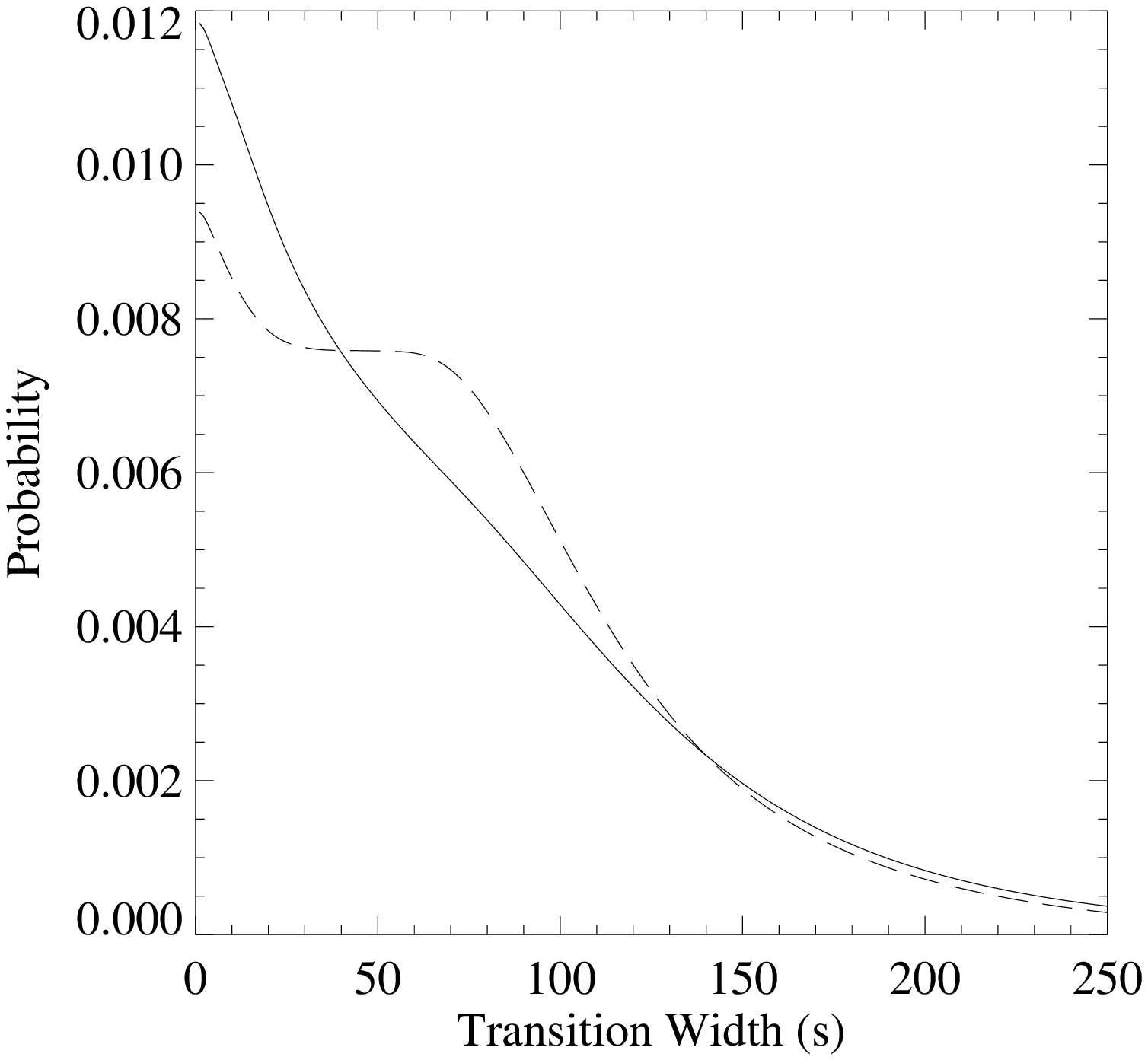}
}
\caption{\small Bayesian probability distribution for the
  width of the eclipse ingress (dashed line) and egress (solid line),
  assuming a linear transition between the eclipsed and uneclipsed
  states.  \protect{\label{fig:egress}}}
\end{figurehere}
\vspace{0.2in}

The fact that the eclipses observed with \chandra\ are total and only
0.08 wide in phase already indicates that the X-ray emission is
emanating from a region nearly two times smaller than in outburst.
The rapidity with which eclipse ingress and egress occurs, however,
can place even tighter constraints.  Assuming that the eclipse is
total and that eclipsing medium has a sharp edge, then the time it
takes for the count rate to go from the mean count rate down to zero,
or the reverse, is indicative of the size of the X-ray source.  As
described in Appendix~\ref{sec:anal}, we have performed a statistical
analysis of the eclipse ingress and egress, and show the results in
Fig.~\ref{fig:egress}.  We have averaged the two ingresses with each
other, and the two egresses with each other, but we have not averaged
the ingresses with the egresses.  At the 95\% confidence level, the
average ingress is more rapid than 170\,sec, while the average egress
is more rapid than 217\,sec.  This implies that the X-ray source
diameter is $\aproxlt 0.07~a$, where $a$ is the binary separation.
For a total system mass of 2\,$\msun$, this implies an X-ray diameter
of $\aproxlt 9\times10^{9}$\,cm (approximately 7 earth diameters).
This is approximately a factor of 15 smaller than the implied X-ray
source diameter for \fu\ in outburst \citep{mcclintock:82a,white:82a}.

An important caveat for the above estimates is that they pertain to
the 0.5--8\,keV lightcurve only, and are dominated by emission from
$\aproxlt 2$\,keV.  Thus, we are unable to distinguish between
different source sizes above and below 2\,keV, as have been
hypothesized \citep{campana:98a,campana:00a}. We also note that for
the most part, each individual ingress and egress looked statistically
identical, with the exception of the second egress.  There was week
evidence for a prolonged period of decreased emission towards the
beginning of the second egress.  This was most readily noted by the
fact that the 1$\sigma$ limit for the second eclipse egress was
260\,sec, as opposed to 130--170\,sec for the other ingresses and
egress.  Accretion ``dips'', due to ``blobs'' of enhanced neutral
hydrogen column are in fact common near zero phase in Cyg~X-1, for
example (see {Ba\-\l{}u\-ci\'nska-Church} et al.  \nocite{balu:97a}
1997 and references therein).  Perhaps a similar occurrence is being
hinted at here, although the statistics are not strong enough to say
with any degree of certainty whether or not this is the case.

The lightcurve not only places constraints upon the X-ray source
diameter, but the measured eclipse duration also places constraints
upon the parameters of the secondary, specifically its mass and
radius\footnote{Previous estimates of these parameters have been
  presented by \citet{horne:86a} utilizing measurements taken during
  outburst of a sinusoidally modulated H$\beta$ absorption line and a
  double peaked {\sc H{\rm e}~ii} 4686 emission line.  The two
  measurements together implied a primary and secondary mass of
  $0.6\pm0.2\,\msun$ and $0.4\pm0.2\,\msun$, respectively.  These
  estimates, however, are hampered by uncertainties in modeling the
  outburst emission of the secondary and disk, as well as by any
  possible contamination from the F star.} Here we calculate the mass
ratio of the secondary to the primary, $q$, as a function of
inclination angle, $i$.  Assuming that the secondary completely fills
its Roche lobe, which we approximate as spherical, then the ratio of
the eclipse duration to the orbital period is given by
\begin{equation}
\theta \equiv \frac{\pi t_{\rm ecl}}{P_{\rm orb}} = \sqrt{ \left ( \frac{R_2}{a} \right
  )^2 - \left ( \frac{\cos i }{1+q} \right )^2 } ~,
\end{equation}
where $a$ is the orbital separation of the binary and $R_2$ is the
Roche lobe radius of the secondary.
Utilizing the approximation of \citet{paczynski:71a}, this can be
rewritten as $q \approx q_0 ( 1 - q_1 )$ where
\begin{eqnarray}
q_0 &\equiv& \frac{81}{8} \left [ \theta^2
   + \cos^2 i \right ]^{3/2} \cr
q_1 &\equiv& \frac{81}{8} \left [ 2 \cos^2 i -  \theta^2
   \right ] \left [ 
  \theta^2 + \cos^2 i \right ]^{1/2} ~.
\end{eqnarray}
Assuming specific values for the primary mass (1.3 and 2.2\,$\msun$),
in Fig.~\ref{fig:massrad} we plot the secondary mass, $M_2$, and Roche
lob radius, $R_2$, as a function of orbital inclination.

\vspace{0.2in}
\begin{figurehere}
  \centerline{\includegraphics[width=0.45\textwidth]{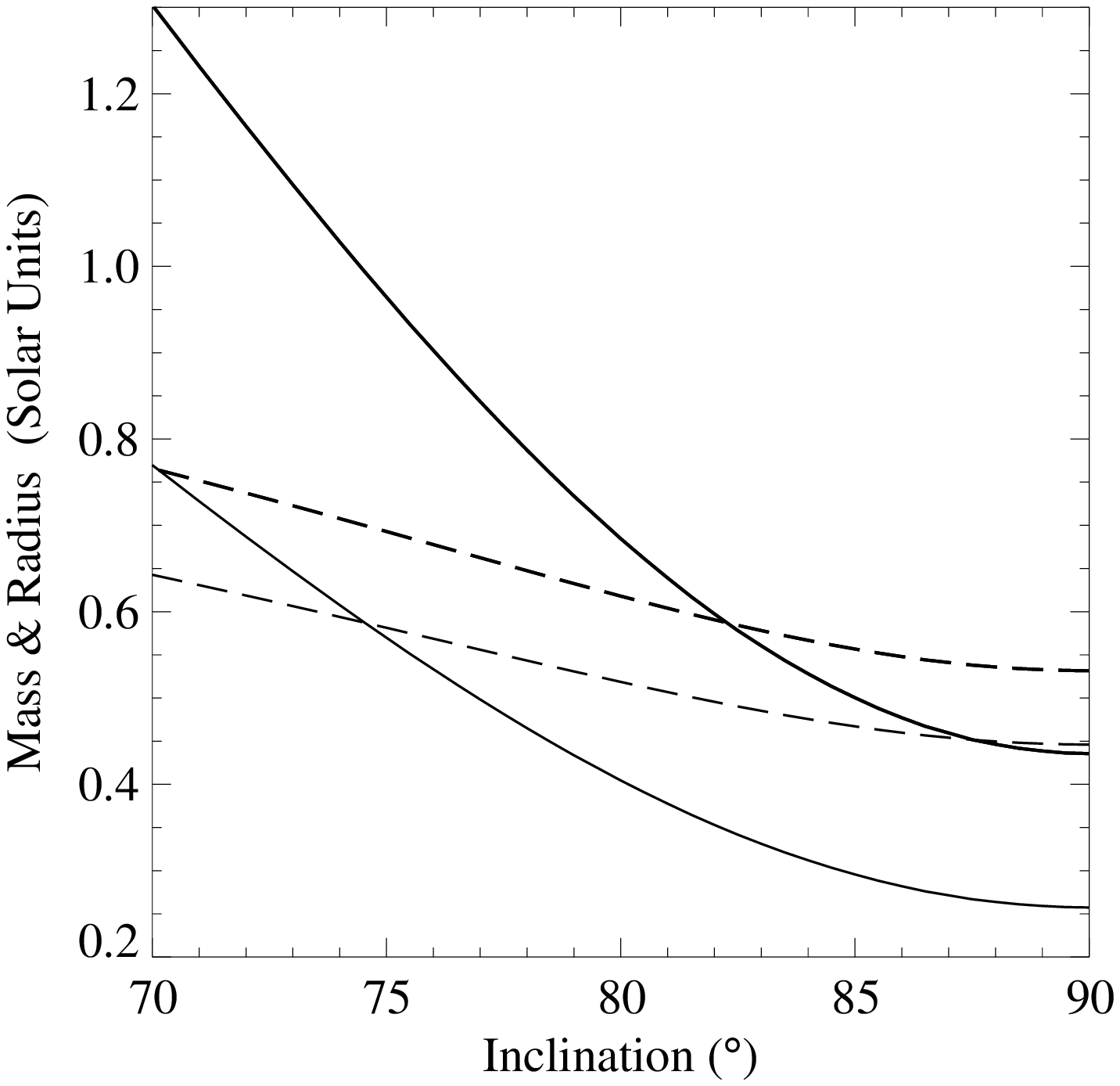} }
\caption{\small  Constraints on the mass (solid lines) and radius (dashed
  lines) of the secondary, as a function of the system inclination.
  We have assumed a primary mass of 1.3\,$\msun$ (lower lines) and
  2.2\,$\msun$ (upper lines).  \protect{\label{fig:massrad}}}
\end{figurehere}
\vspace{0.2in}

If one takes as a straw-man model that the primary is a neutron star
with mass 1.3--2.2\,$\msun$ (as evidenced by the Type I burst observed
by Garcia \& Grindlay \nocite{garcia:87a} 1987) and that the Roche
lobe radius of the secondary is greater than or equal to its main
sequence radius (as has been measured for the secondary in numerous
accreting systems), then the eclipse measurements alone imply that the
system inclination $i > 75^\circ$ and that for the secondary mass
$0.25\,\msun < M_2 < 0.6\,\msun$ (i.e., $0.2 \aproxlt q \aproxlt
0.5$).

The other interesting aspect to note about the observed lightcurve is
that the position of the (out of eclipse) lightcurve minimum occurs at
a completely different phase than in outburst.  In the lightcurve
presented by \citet{mcclintock:82a}, the (weak) non-eclipse minimum
occurred at phase $\approx 0.8$.  That is, the dip leads the eclipse by
$\approx 0.2$ in phase, similar to what is found for X1822$-$371
\citep{parmar:00a,heinz:01a}.  In quiescence, the non-eclipse minimum
occurs at phase $\approx 0.4$, i.e., it leads the eclipse by 0.6 in
phase.  In a model where the minimum corresponds to the location of
where the accretion stream first interacts with the accretion disk,
this shift can be explained if the disk has substantially shrunk in
size.  

As a rough illustration of this, consider a crude model
where the incoming matter has the specific angular momentum solely due
to orbital motion at the inner Lagrange point, and otherwise is only
falling in the gravitational potential of the compact object (i.e., we
are assuming $q \ll 1$).  Assuming an elliptic orbit intersecting the disk, 
it is relatively straightforward to show that
the radius of the accretion disk, $R_D$, relative to the binary
separation is given by
\begin{equation}
\frac{R_D}{a} = \frac{x^3}{\left[ (x^3 - 1) \cos \phi + 1 \right ]} ~~,
\end{equation}
where $x \equiv (a - R_2)/a$, $R_2$ is the Roche lobe radius of the
secondary, and $\phi$ is the phase (in radians) by which the dip leads
the eclipse.  (The fact that here $\phi > \pi$ indicates that the
approximations are not fully valid; however, they are sufficient to
illustrate the general point.)  Approximating the size of the Roche
lobe as a function of mass ratio as above, one finds, for a broad
range of $q$, that $R_D \approx 0.4$--$0.5\,a$ for $\phi=0.2$.  This
is consistent with independent estimates of the disk radius of
X1822$-$371 \citep{heinz:01a}.  Taking $\phi = 0.5$ and $q = 0.2$, one
finds the substantially smaller value of $R_D \approx 0.2\,a$.  More
accurate estimates of the disk radius would rely upon obtaining better
estimates of the binary inclination and mass ratio, $q$.

\section{Spectrum}\label{sec:spec}

In Fig.~\ref{fig:piled} we present the 0.5--8\,keV spectrum of \fu,
averaged over all phases except the eclipse.  Note that in this figure
we have not corrected for photon pile-up, although we do account for
this effect in our fits.  Overall, the 0.5--2\,keV spectrum is very
similar to the \rosat\ \textsl{Position Sensitive Proportional
  Counter} (\pspc) spectrum described by \citet{garcia:99a}. The
\chandra\ spectrum, however, clearly shows the presence of a hard
tail, even after accounting for pile-up, that lies beyond the upper
energy cutoff of \rosat.

As a first description of the \chandra\ spectrum, we fit a model
consisting of a blackbody plus power-law that are absorbed by a
neutral column (utilizing the absorption model of Wilms, Allen, \&
McCray \nocite{wilms:00a} 2000).  Results of this fit are presented in
Table~\ref{tab:fit} (model A), where we have parameterized the
blackbody normalization by an effective radius, $R$, equivalent to
assuming an emitting area of $\pi R^2$ and a source distance of
6\,kpc.  The power-law flux is for the photon index being $\propto
E^{-\Gamma}$, and the power law normalization is photons ${\rm
  keV^{-1}\,cm^{-2}\,s^{-1}}$ at 1\,keV.  (Note the large error bars
on this normalization are more indicative of the uncertainties in the
power law slope, as opposed to the flux uncertainties at energies
$>2$\,keV.)  The neutral column, blackbody temperature, and implied
flux we obtain are consistent with the fits to \pspc\ observations
described by \citet{garcia:99a}, which were also averaged over phase.
The flux and spectrum of \fu\ is therefore consistent with having
remained constant since 1994.\footnote{Earlier \rosat\ \textsl{High
    Resolution Imager} (\hri) observations \citep{garcia:94a} lacked
  any detailed spectral information.  However, assuming the same model
  as for the \pspc\ fits, \citet{garcia:99a} argued that during the
  1992 \hri\ observations, \fu\ was approximately a factor of two more
  luminous.  \citet{rutledge:00a}, using neutron star atmosphere
  models, also argued that \fu\ was likely more luminous during the
  \hri\ observations.}

\vspace{0.2in}
\begin{figurehere}
\centerline{
\includegraphics[width=0.45\textwidth]{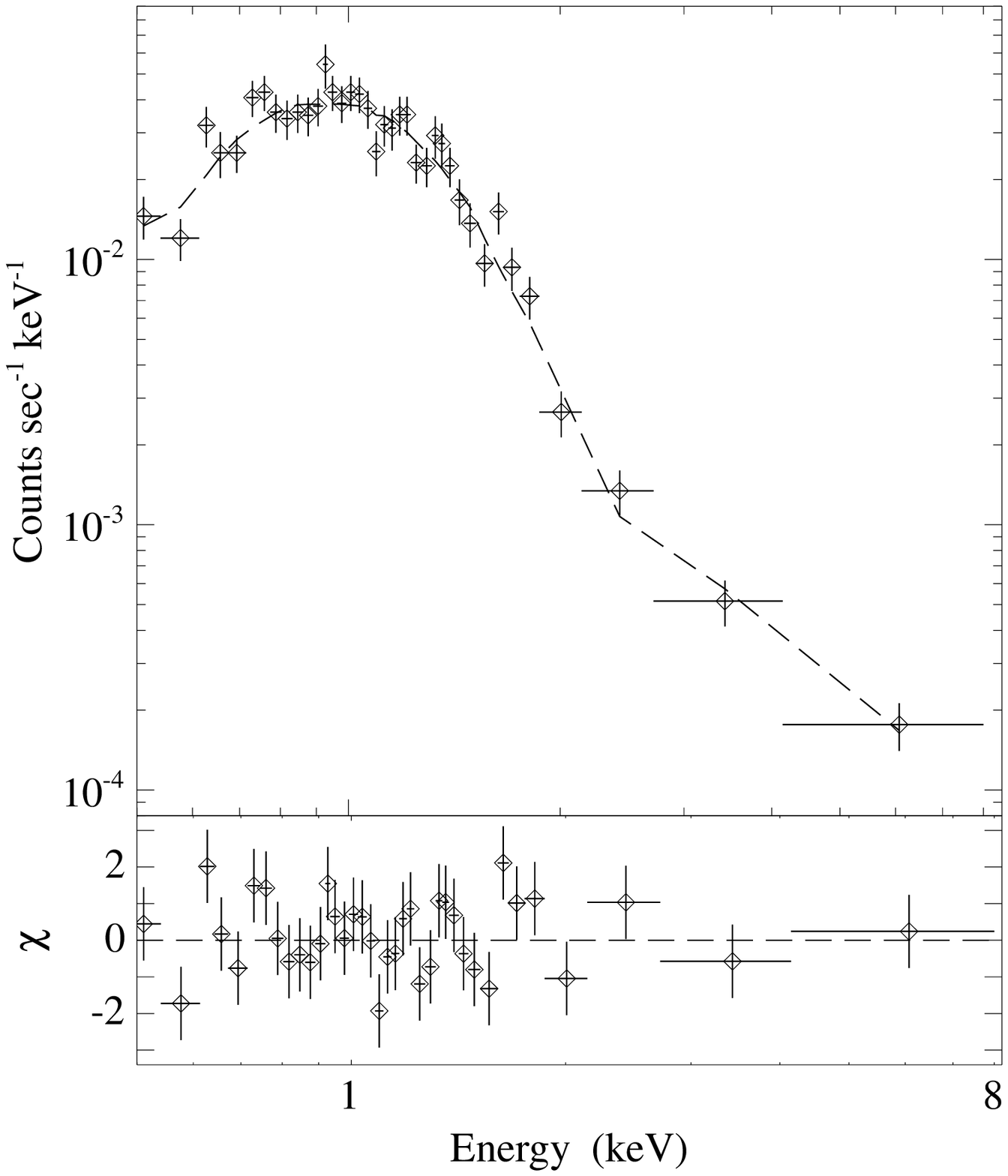}
}
\caption{\small Measured 0.5-8\,keV spectrum of \fu, \emph{excluding} the
  eclipses, fitted with an absorbed blackbody plus power law.  Lower
  panel shows the residuals from this fit.
  \protect{\label{fig:piled}}}
\end{figurehere}
\vspace{0.2in}

The fitted radius of $< 4$\,km (90\% confidence levels) for these
observations is significantly smaller than the $\approx10$\,km
expected for a neutron star; however, as discussed by
\citet{rutledge:00a}, this might be attributed to the fact that a
simple blackbody is not the correct model for a realistic neutron star
atmosphere.  Numerous theoretical models exist that more carefully
treat the radiative transfer in such an atmosphere. To describe the
\chandra\ spectra of \fu, we here use an {\tt ISIS}
\citep{houck:00a} implementation of the model of \citet{zavlin:96a}.
Fit parameters for this model include neutron star temperature, mass,
radius, and distance.  For all fits described here, we have fixed the
neutron star mass to 1.4\,$\msun$.  As presented in
Table~\ref{tab:fit}, we have performed fits with a fixed distance of
6\,kpc (model C), and fixed radii of 10\,km (model E) and 5\,km (model
G).  The parameter space for these fits contain a number of local
$\chi^2$ minima, and the ``best fit'' parameters and their associated
error bars describe the statistics within the parameter regions of
these local minima.

We see that fixing the distance at 6\,kpc leads to good fits with a
slightly larger neutron star radius of $\aproxgt 5$\,km and a
temperature very similar to that of the simple blackbody fit.
Postulating a larger neutron star radius of 10\,km yields a lower
temperature of $\approx 0.1$\,keV--- consistent with that found by
\citet{rutledge:00a} using these same models to describe the \pspc\ 
data; however, the inferred distance is then rather small.  The 90\%
confidence upper limit of $<4$\,kpc is less than most of the distance
estimates based upon optical studies of the F star.  (Note that the
fit with the fixed 10\,km radius also yielded the largest fitted value
of the neutral column.)  Assuming instead a neutron star radius of
5\,km yields distance estimates consistent with those of the F star.

\vspace{0.2in}
\begin{figurehere}
\centerline{
\includegraphics[width=0.45\textwidth]{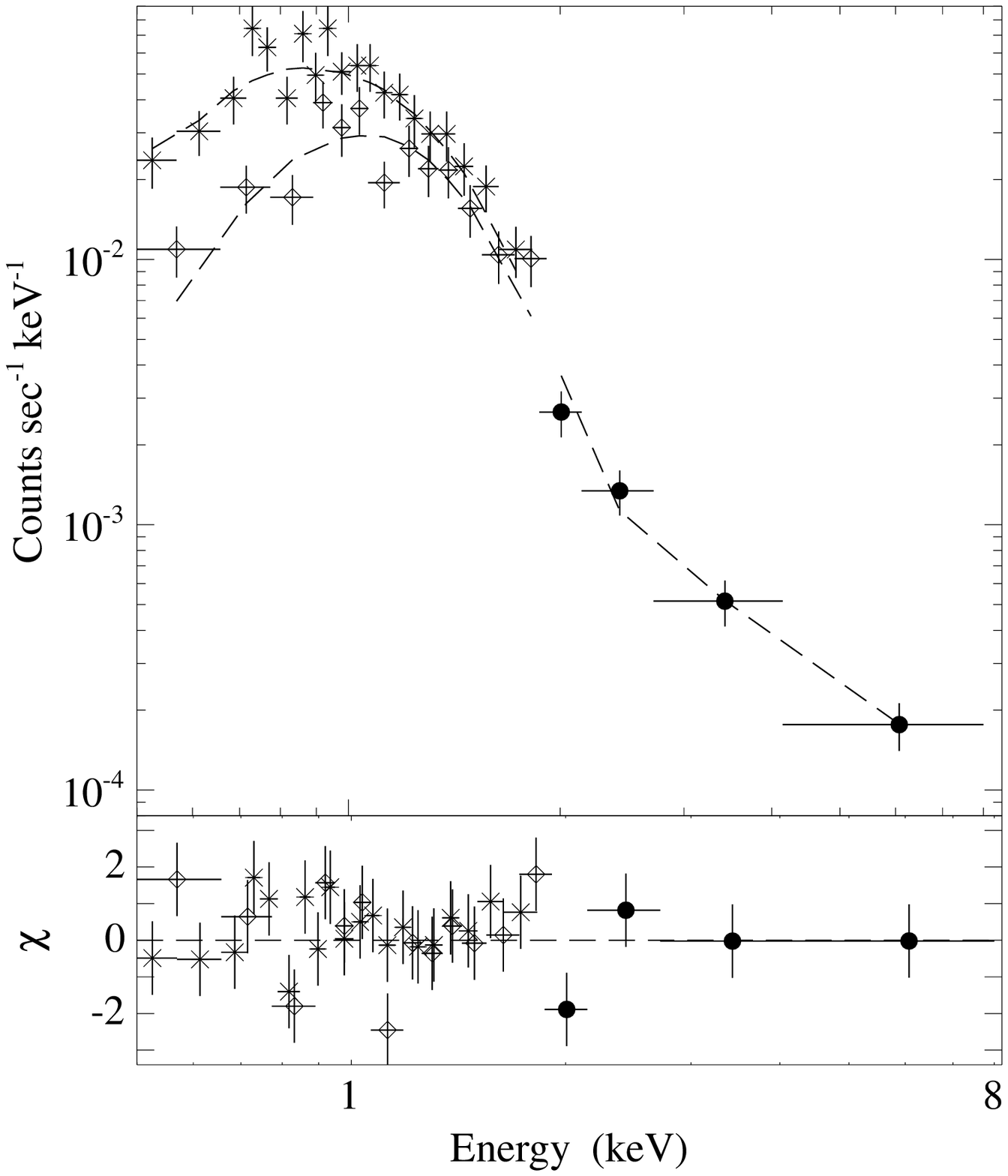}
}
\caption{\small Phase resolved spectra of \fu\ (upper panel).  The filled
  circles represent the 2--8\,keV spectrum, measured from all phases
  excluding the eclipses.  Diamonds and asterisks represent the
  spectrum integrated over phases 0.65--0.96 and 0.2-0.5,
  respectively.  Dashed line is the best fit absorbed blackbody plus
  power law model where all parameters, except neutral hydrogen
  column, were constrained to be identical as a function of phase.
  Lower panel shows the contribution to the $\chi^2$ of the residuals
  from this fit.  \protect{\label{fig:phased}}}
\end{figurehere}
\vspace{0.2in}

It had already been noted by \citet{garcia:99a} that the neutral
column obtained from the orbital phase-averaged \rosat\ \pspc\ 
spectrum was too large to be consistent with the reddening of the F
star.  Using our typical fitted column of $4\times10^{21}~{\rm
  cm^{-2}}$ implies E(B$-$V)$\approx0.75$ \citep{predhel:95a}, whereas
the E(B$-$V)$=0.34$ measured by \citet{deutsch:96a} implies a column
of $1.9\times10^{21}~{\rm cm^{-2}}$.  Phase resolved spectroscopy
offers a possible solution to this discrepancy.  As shown in
Fig.~\ref{fig:lcurve}, there is clear phase-dependent variability.
Extracting spectra at orbital phases 0.2--0.5 and 0.65--0.958, the
variation appears to be consistent with changes in the soft spectrum
only.  (The statistics of the hard tail above 2\,keV become poor in
each phase bin, but are consistent with a constant amplitude and
slope.)  We therefore have chosen to perform a series of fits where
below 2\,keV we consider the above two orbital phases, while above
2\,keV we use the spectrum from the total observation, excluding the
eclipse.  We constrain all model parameters, including normalizations,
to be the same for each of these spectra, except that we allow the
neutral column to be different in each orbital phase.  (For the hard
tail, we constrain the neutral column to be the same as for orbital
phases 0.65--0.958.)  All three spectra are fitted simultaneously, and
the results are presented in Table~\ref{tab:fit} and
Fig.~\ref{fig:phased}.  As for our orbital phase-averaged spectra, we
present a simple blackbody model (model B), and neutron star
atmosphere models with fixed distances (model D) or radii (models F,
H).

Overall these fits are fairly good, and we see that the fitted
temperatures, radii, distances, and power law parameters are very
comparable to those obtained for the phase-averaged fits.  The neutral
column is seen to vary by nearly a factor of two between the two
phases, with the lower column value being a factor of only 10-70\%
larger than would be implied by the F star reddening.  The averages of
the two column values are approximately the same as the columns
determined from the phase-averaged fits.  The spectral variations are
therefore consistent with being solely due to variations of the
intrinsic column, with the total column having roughly equal
contributions from intrinsic and extrinsic columns.

As regards the power law components in these fits, all fits imply that
the power law comprises $\aproxgt 10\%$ of the (unabsorbed)
0.5--8\,keV flux.  The error bars are large; however, the index of the
power law is consistent with being fairly flat.  We note that, if
anything, changing our assumptions regarding pile-up is most likely to
further flatten the power law (see \S\ref{sec:extract}). In terms of
amplitude, slope, and extent, such a power law is difficult to
describe solely using a neutron star atmosphere model.

\section{Discussion}\label{sec:summary}

Our \chandra\ observations of \fu\ have revealed that the previously
observed extended coronal structure has collapsed by at least a factor
of 15 in size.  The soft X-ray spectrum is now consistent with thermal
emission from the surface of a neutron star with radius on the order
of 5\,km.  To what extent this emission is indicative of residual
thermal emission or active, albeit low-level, accretion, remains
unclear. The fact that the spectrum observed by \chandra\ is
consistent in amplitude and shape with that observed by \rosat\ in
1994 can easily be accommodated for in the cooling neutron star model,
as the cooling time scale is expected to be very long.  The prior
observed soft X-ray variability \citep{garcia:94a} can be accounted
for via orbital phase-dependent obscuration effects, rather than
intrinsic accretion variability.

On the other hand, questions concerning the cooling model remain.
First, assuming a distance of 6\,kpc, consistent with various
estimates of the F star distance, the fitted neutron star radius tends
to be be fairly low, on the order of 5\,km.  Although larger than the
inferred blackbody radius, this is smaller than expectations from
numerous neutron star models.  Forcing a large neutron star radius,
10\,km, implies a distance substantially less than most estimates of
the F star distance, which of course might still be a chance overlap.
However, it is seen that each of the various fits, including the
simple blackbody, are statistically nearly identical to one another.
Second, the variability discussed by \citet{garcia:94a} is consistent,
in amplitude and phase, with that observed here, which indicates that
roughly 2/3 of the integration time from the 1992 \hri\ observation
occurred over the phases of \emph{lowest} emission.  Thus the estimate
of an approximate factor of two flux difference between 1992 and 1994
might be an \emph{underestimate}, although this is hampered by the
poor spectral resolution of the \hri.  \citet{rutledge:00a} argue that
this change might be related to solely the outer crust cooling.
Additionally, this change in flux might indicate that as of 1992, \fu\ 
had not yet truly entered into a quiescent phase.

Based upon the work of \citet{brown:98a} and the observations
discussed by \citet{garcia:94a}, \citet{garcia:99a}, and
\citet{rutledge:00a}, \citet{wijnands:02a} has suggested that the
quiescent phase of \fu\ might be too faint to be consistent with the
neutron star cooling model.  Specifically, one expects a quiescent
flux of
\begin{equation}
F_q \approx \frac{t_o}{t_o+t_q} \frac{\langle F_o \rangle}{135} ~~,
\end{equation}
where $\langle F_o \rangle$ is the average flux in outburst, $t_o$ is
the average time the source spends in outburst, and $t_q$ is the
average time spent in quiescence.  Assuming a typical accretion
efficiency of $\epsilon \sim 0.1$, and an average accretion rate of
$\langle \dot M \rangle \sim 4 \times 10^{18}~{\rm g~s^{-1}}$, as
suggested by the possible period derivative found in
\S\ref{sec:curve}, we then find that
\begin{equation}
\frac{t_q}{t_o} \aproxgt 10^3 \left ( \frac{D}{6~{\rm kpc}} \right )^2 
     \left( \frac{\epsilon}{0.1} \right ) ~~.
\end{equation}
To accommodate the above result, either the assumed accretion rate is
atypically large (i.e., not indicative of the long term average in
outburst), the durations of the quiescent periods are very long, there
is enhanced cooling of the neutron star during quiescence (see
Wijnands \nocite{wijnands:02a} 2002, Wijnands et al.
\nocite{wijnands:02b} 2002, and references therein), or the efficiency
is very low.

If one instead assumes an average outburst flux of $\langle F_q
\rangle \sim 2 \times 10^{-9}~{\rm erg~cm^{-2}~s^{-1}}$ (i.e., 5 times
the typical 2--10\,keV flux seen in outburst; \citealt{garcia:87a} and
references therein), then one obtains $t_q/t_o \aproxgt 20$.  However,
this estimate assumes that one has observed nearly all of the outburst
flux in the system, contrary to the more usual assumptions for an ADC
of either a very optically thin or very optically thick coronae seen
near edge on \citep{parmar:00a,heinz:01a}.  In order to retain the
typical ADC assumptions, one is again drawn toward hypotheses of the
previous outburst episode being atypically bright, long quiescent
periods, enhanced neutron star cooling, or very low radiative
efficiencies.  Given that any matter that reaches the neutron star
surface will contribute towards heating the neutron star, the latter
hypothesis implies that inefficiency during outburst is achieved via
an outflow, e.g., an ADIOS.

Questions remain about the origin of the sinusoidal variability and
the power law tail.  The former is similar to what has been observed
in X1822$-$371, where the variability has been associated with the
interaction of the accretion stream with the outer edge of an
accretion disk \citep{parmar:00a,heinz:01a}.  A similar model for \fu\ 
could argue for active accretion.  Observations of X1822$-$371 imply
that a totally opaque disk rim is raised that completely blocks the
central neutron star, while a more vertically extended disk atmosphere
(with $N_{\rm H} \aproxlt 10^{22}~{\rm cm^{-2}}$) attenuates the
extended corona \citep{heinz:01a}.  Here the evidence suggests that
the neutron star in \fu\ is directly observable, while the combination
of inclination angle and outer disk atmosphere height serves only to
attenuate its soft flux with a column of $N_{\rm H} \approx 2 \times
10^{21}~{\rm cm^{-2}}$.  The presence of the hard tail might also
argue for active accretion.  The limited \chandra\ band pass shows
that at least $\aproxgt 10\%$ of the total, unabsorbed 0.5--8\,keV
luminosity emanates from this component.  Thus we might be seeing the
accretion energy first being funneled through a corona, and then
actively being dissipated as thermal energy on the neutron star
surface.

One alternative that preserves the cooling model would be for the
accretion stream to be halted by an outgoing pulsar wind
\citep{campana:98a,menou:99a,campana:00a}.  The hard tail would
represent shocked emission from the interaction region of the wind and
accretion stream.  The sinusoidal modulation could still be due a
neutral hydrogen column being raised by the interaction of the
accretion stream with a disk rim. Such a disk need not be actively
accreting, however, as the pulsar wind and magnetic field could halt
any inward flow of disk material.  We note that the disk radius
inferred from the location of the light curve minimum is comparable to
expectations from the pulsar wind model; however, the soft flux radius
determined from the rapidity of the eclipse is somewhat smaller.

The \chandra\ observations have lent further credence to the
hypothesis that the \fu\ system is a hierarchical triple.  The X-ray
source position is now seen to be even more closely aligned with
optical position of the F star.  Furthermore, the minimum fitted
neutral column (i.e., the flux from the maximum emission phase of the
lightcurve) is more consistent with the reddening of the F star than
were previous measurements obtained from phase-averaged spectra.

Prospects are good for refining the answers to some of the remaining
questions about \fu.  Further observations with either \chandra\ or
\xmm, aside from improving the statistics of the model fits, might
reveal variability.  Variability in the soft spectra would argue
against the cooling model be the sole source of quiescent emission.
\xmm\ observations can have direct bearing upon the pulsar wind models
for the hard tail.  Aside from extending detection of this tail to
beyond 8\,keV (which would rule out most atmosphere models for the
hard tail), the methods of \S\ref{sec:curve} then could be applied
separately to the hard tail.  Measurement of three or more eclipses
would allow one to place limits of $\approx 0.1~a$ on the diameter of
the hard component emission region.  Although large, this is a smaller
diameter than suggested by many of the pulsar wind models.

Additionally, each \xmm\ observation can determine the eclipse
centroid to approximately $\pm 8$\,sec.  If the F star is truly in a
month long orbit about the binary, then we expect $\pm 80$\,sec
variations in the eclipse ephemeris.  Thus, multiple \xmm\ 
observations spaced out over a month long time scale could directly
determine the orbital parameters of a third body.  In the absence of
the detection of such orbital variations, either \chandra\ or \xmm\ 
observations could determine whether or not the orbital period
evolution hinted at in the analysis of \S\ref{sec:curve} actually
occurred.  This in turn, as discussed above, could have implications
for the expected level of quiescent emission in the cooling neutron
star model.

\acknowledgements We would like to acknowledge C.  Bailyn, L.
Bildsten, J.  Davis, J. McClintock, and T. Maccarone for useful
conversations, and P.  Maloney for providing a moment of Zen.  M.A.N.
would like to acknowledge the hospitality of the Yale Astronomy
Department, where much of this work was carried out.  We thank V.
Zavlin for providing us with the 
neutron star atmosphere code.  Support for this work was provided by
the National Aeronautics and Space Administration through Chandra
Award Number G00-2038X issued by the Chandra X-ray Observatory Center,
which is operated by the Smithsonian Astrophysical Observatory for and
on behalf of NASA under contract NAS8-39073.

\appendix

\section{Eclipse Analysis}\label{sec:anal}

In what follows, we explicitly ignore the effects of detector ``deadtime''.
Each \axaf\ frame for this observation represented an integration time of
approximately 1.1\,s with a ``frame transfer time'' of 40\,ms between
frames\footnote{See the \chandra\ Proposer's Observatory Guide,
  http://cxc/harvard.edu.}.  The effective deadtime therefore was on the
order of 3.5\%, which is negligible for the purposes of this analysis.  We
shall employ a Bayesian approach.

The following discussion pertains to eclipse egress; however, it is readily
generalized to ingress as well.  We label our $\Delta t =1.1$\,s time bins
with indices $0 \rightarrow k$, and assume that at the beginning of the
time bin indexed by 0, the X-ray source is fully eclipsed, and by the
beginning of the time bin indexed by $k$ it is fully unobscured.  We assume
that the mean count rate measured at the detector increases linearly from 0
to $r$ (although other assumptions are readily accommodated).  The mean
count rate in time bin $i$ is therefore $r_i = r (i+1/2)/k$, for $i = 0
\rightarrow k-1$.  Following the usual derivation of Poisson statistics, we
assume that each bin is short enough such that the probability of measuring
one photon event in a given bin is $P(1) = r_i \Delta t$, the probability
of measuring no events is $P(0) = 1 - P(1)$, and that $P(>1) =0$.  The
probability of the data, $D$, given this model is therefore
\begin{equation}
P(D|r,k) = \prod_i r_i \Delta t \prod_j (1 - r_j \Delta t) ~~,
\end{equation}
where the product over $i$ is for those bins where a photon was detected,
and the product over $j$ is for those bins where no photons were detected.
We divide the above by the probability assuming a steady,
uneclipsed count rate of $r$, which, given $r \ll 1$,  yields
\begin{equation}
P(D|r,k) \propto \prod_i \left ( \frac{r_i}{r} \right ) 
                 \prod_j [1 + (r-r_j) \Delta t] 
 =  \prod_i \left ( \frac{i + 1/2}{k} \right ) 
                 \prod_j \left [1 + r \Delta t \left ( 
                            \frac{k-j-1/2}{k} \right ) \right ] ~~.
\end{equation}

Further approximating the product over empty time bins as the exponential
of a sum, taking a uniform prior for $k$ (the eclipse sharpness) and a
gaussian prior for the detected count rate, and employing Bayes theorem, we
obtain the probability distribution for the model parameters as
\begin{equation}
P(r,k|D) \propto \exp \left [ {-(r - \bar r)^2}/{2 \sigma_r^2} \right ]
                 \times 
\prod_i \left ( \frac{i + 1/2}{k} \right ) 
                 \exp \left [r \Delta t \sum_j \left ( 
                            \frac{k-j-1/2}{k} \right ) \right ] ~~.
\label{eq:pdak}
\end{equation}
In practice, based upon the fits to the folded lightcurve, we have used
$\bar r = 0.046$\,cps and 0.042\,cps for the mean ingress and egress count
rates, respectively, and we have taken $\sigma_r = 0.0035$\,cps for both.
In the above we have assumed a fixed start point for the egress
commencement.  We allow for uncertainty in this start point by assuming
that its prior probability is uniformly distributed between 0 and 500\,s
before the first detected photon. Eq.~(\ref{eq:pdak}) then must be slightly
modified by including in the sum within the exponential all the time bins
from 500\,s before the first detected photon to the time bin represented by
index 0.
                 
The above was used to calculate probability distributions separately for
the two measured eclipse ingresses and the two measured eclipse egresses.
The probability distributions were then multiplied together for each pair
to give the ingress and egress distributions for the folded lightcurve.
Marginalizing over the eclipse sharpness yielded the eclipse widths and
midpoints, while marginalizing over the eclipse start points yielded the
distributions for the eclipse sharpness (see \S\ref{sec:curve}).

\section{Serendipitously Observed Sources}\label{sec:serendipity}

A number of other sources were serendipitously observed within a few
arcminute radius of \fu.  These sources, along with their coordinates
and 0.3--8\,keV total counts (i.e., integrated over the entire
36.6\,ksec observation) are presented in Table~\ref{tab:serendipity}.
For the most part, these sources can be identified with stellar
objects in the Digital Sky
Survey\footnote{http://www-gsss.stsci.edu/DSS/dss\_home.htm}.  Several
sources, specifically S3-$\beta$, S2-$\alpha$, I2-$\alpha$, and
I3-$\alpha$ might be faint active galactic nuclei.  S3-$\beta$
contains sufficient counts to fit a crude spectrum, and it is found to
be consistent with an absorbed power law, with $N_{\rm H} = 3\pm3
\times 10^{21}~{\rm cm^{-2}}$ and $\Gamma=1.9\pm0.6$, 90\% confidence
limits.  Its 0.5--8\,keV (absorbed) flux corresponds to
$4\times10^{-14}~{\rm erg~cm^{-2}~s^{-1}}$.  There is also weak
evidence that its count rate increases by a factor of two on time
scales of $\approx 25$\,ksec.

The source S3-$\delta$ was useful for performing differential
astrometry, as it likely was detected also with the \hst\ observations
of \citet{deutsch:96a}.  For purposes of differential astrometry (as
opposed to Table~\ref{tab:serendipity}), the source position was
determined with the program {\tt wavdetect} applied to the 0.5--2\,keV
image.

Note that given the broad energy range (0.3--8\,keV), and large
confidence contours (3-$\sigma$ detection region from {\tt
  wavdetect}), the total counts in Table~\ref{tab:serendipity} might
contain photons unrelated to the source, and therefore should be
regarded as upper limits.  This is especially true as the angular
distance from the target point, S3-$\alpha$, increases.


\clearpage

\begin{deluxetable}{lcccl}
\tablewidth{0pt}
\tablecaption{Optical identifications for the quiescent counterpart to the 
  \fu\ system. \label{tab:optical}}
\tablehead{ \colhead{Wavelength} & \colhead{E(B$-$V)} & \colhead{Dist.}
      & \colhead{Spec. Type} & \colhead{Reference}}
\startdata
4500--7500\,\AA & 0.3 & $>1$\,kpc & & Thorstensen et
al. \nocite{thorstensen:88a} 1988 \\
4800--6600\,\AA & 0.5 & 4\,kpc & F5-8\,V & Chevalier et
al. \nocite{chevalier:89a} 1989 \\
4000--4800\,\AA & & & F7\,IV-V & Garcia et al. \nocite{garcia:89a} 1989 \\
3700--5600\,\AA & 0.3 & 6--8\,kpc & F8\,IV & Cowley \& Schmidtke
\nocite{cowley:90a} 1990 \\
2200-3300\,\AA & 0.34 & &  F8\,IV & Deutsch et al. \nocite{deutsch:96a} 1996 \\
\enddata
\end{deluxetable}

\begin{deluxetable}{lccccccccr}
\tablewidth{0pt}
\tablecaption{Fits to the X-ray Spectra of \fu\ (90\% Confidence Level 
       error bars). \label{tab:fit}}
\tablehead{ \colhead{Fit} & \colhead{${\rm N_H}$} & \colhead{kT} & 
     \colhead{R or D \tablenotemark{a}} & \colhead{$A_\Gamma$} & \colhead{$\Gamma$}
     & \colhead{0.5-2\,keV \tablenotemark{b}} 
     & \colhead{0.5-2\,keV \tablenotemark{c}} 
     & \colhead{2-8\,keV} & \colhead{$\chi^2$/DoF} \\
& \colhead{$(10^{21}\,{\rm cm}^{-2})$} & \colhead{(keV)} & \colhead{(km or kpc)}
     & \colhead{($10^{-6}$)} 
     & & \multicolumn{3}{c}{$(10^{-13}\,{\rm erg~cm^{-2}~s^{-1}})$} }
\startdata
A & \errtwo{3.1}{0.9}{0.7} & \errtwo{0.21}{0.02}{0.03} 
& \errtwo{2.5}{1.5}{0.7} & \errtwo{6.4}{20.2}{5.2}
& \errtwo{1.1}{1.1}{1.1}
& 1.0 & 2.5 & 0.5 & 38.1/33 \\
B & \errtwo{2.0}{0.9}{0.8} & \errtwo{0.22}{0.03}{0.02} 
& \errtwo{2.1}{1.0}{0.7} & \errtwo{2.8}{10.0}{2.4}
& \errtwo{0.6}{1.1}{1.3}
& 1.3 & 2.5 & 0.5 & 37.3/31 \\
\nodata & \errtwo{4.3}{1.1}{1.0} & \nodata 
& \nodata  & \nodata 
& \nodata 
& 0.8 & \nodata & \nodata  & \nodata \\
C & \errtwo{3.4}{0.3}{0.4} & \errtwo{0.25}{0.00}{0.06} 
& \tablenotemark{d} $5.0^{+0.9}$ & \errtwo{2.1}{0.5}{1.8}
& \errtwo{0.4}{0.6}{1.1}
& 1.0 & 2.9 & 0.5 & 38.2/33 \\
D & \errtwo{2.7}{0.4}{0.7} & \errtwo{0.24}{0.01}{0.13} 
& \tablenotemark{d} $5.0^{+2.4}$ & \errtwo{1.5}{5.2}{1.3}
& \errtwo{0.2}{1.1}{1.4}
& 1.3 & 3.1 & 0.5 & 38.0/31 \\
\nodata & \errtwo{5.0}{0.6}{1.0} & \nodata 
& \nodata  & \nodata 
& \nodata 
& 0.8 & \nodata & \nodata  & \nodata \\
E & \errtwo{4.2}{1.2}{0.9} & \errtwo{0.08}{0.03}{0.02} 
& \errtwo{1.9}{2.1}{1.2} & \errtwo{4.6}{20.1}{3.9}
& \errtwo{0.9}{1.3}{1.4}
& 1.0 & 4.0 & 0.5 & 37.6/33 \\
F & \errtwo{3.2}{1.1}{1.1} & \errtwo{0.10}{0.02}{0.01} 
& \errtwo{2.7}{3.3}{1.6} & \errtwo{2.2}{8.1}{2.0}
& \errtwo{0.4}{1.2}{1.4}
& 1.3 & 4.0 & 0.5 & 37.9/31 \\
\nodata & \errtwo{5.4}{1.4}{1.2} & \nodata 
& \nodata  & \nodata 
& \nodata 
& 0.8 & \nodata & \nodata  & \nodata \\
G & \errtwo{3.9}{1.0}{0.9} & \errtwo{0.22}{0.05}{0.05} 
& \errtwo{4.2}{3.6}{2.2} & \errtwo{3.7}{15.8}{3.3}
& \errtwo{0.8}{1.2}{1.3}
& 1.0 & 3.6 & 0.5 & 37.4/33 \\
H & \errtwo{2.8}{1.1}{0.8} & \errtwo{0.25}{0.05}{0.05} 
& \errtwo{5.8}{4.2}{3.0} & \errtwo{1.6}{8.2}{1.5}
& \errtwo{0.3}{1.2}{1.6}
& 1.3 & 3.6 & 0.5 & 38.0/31 \\
\nodata & \errtwo{5.1}{1.3}{1.1} & \nodata 
& \nodata  & \nodata 
& \nodata 
& 0.8 & \nodata & \nodata  & \nodata \\
\enddata
\tablenotetext{a}{Neutron star radius (km) assuming a distance of 6\,kpc
  (A--D), or neutron star distance (kpc) assuming a radius of 10\,km (E--F)
  or 5\,km (G--H).}
\tablenotetext{b}{Absorbed Flux.} \tablenotetext{c}{Unabsorbed Flux.}
\tablenotetext{d}{Lower limit set by table boundaries of neutron star
  atmosphere model.} 
\end{deluxetable}

\begin{deluxetable}{lccr}
  \tablewidth{0pt} \tablecaption{Sources in the \chandra\ 
    field. \label{tab:serendipity}} \tablehead{
    \colhead{Source\tablenotemark{a}} & \colhead{RA} & \colhead{DEC} &
    \colhead{Counts\tablenotemark{b}} \\
    & \multicolumn{2}{c}{(J2000)} & (0.3--8\,keV)} \startdata S3-$\alpha$ &
  ${\rm 21^h\,31^m\,26.19^s}$ & $47^\circ\,17\arcmin\,24.7\arcsec$
  & 1301   \\
  S3-$\beta$ & ${\rm 21^h\,31^m\,24.57^s}$ &
  $47^\circ\,17\arcmin\,12.7\arcsec$
  & 205   \\
  S3-$\gamma$ & ${\rm 21^h\,31^m\,30.29^s}$ &
  $47^\circ\,16\arcmin\,16.5\arcsec$
  & 36   \\
  S3-$\delta$ & ${\rm 21^h\,31^m\,27.86^s}$ &
  $47^\circ\,17\arcmin\,32.5\arcsec$
  & 28   \\
  S3-$\epsilon$ & ${\rm 21^h\,31^m\,29.21^s}$ &
  $47^\circ\,16\arcmin\,36.9\arcsec$
  & 25   \\
  S2-$\alpha$ & ${\rm 21^h\,31^m\,34.64^s}$ &
  $47^\circ\,15\arcmin\,3.6\arcsec$
  & 88 \\
  S2-$\beta$ & ${\rm 21^h\,31^m\,33.18^s}$ &
  $47^\circ\,15\arcmin\,4.3\arcsec$
  & 9  \\
  I2-$\alpha$ & ${\rm 21^h\,30^m\,55.13^s}$ &
  $47^\circ\,09\arcmin\,49.6\arcsec$
  & 115  \\
  I3-$\alpha$ & ${\rm 21^h\,30^m\,27.83^s}$ &
  $47^\circ\,15\arcmin\,23.5\arcsec$
  & 80  \\
  \enddata \tablenotetext{a}{Chip ID-source identifier (brightest to
    faintest).}  \tablenotetext{b}{Counts from within the 3-$\sigma$
    detection region as determined via {\tt wavdetect}.}
\end{deluxetable}
\end{document}